\documentclass[journal]{IEEEtran}
\usepackage{amsmath}
\usepackage{breqn}
\usepackage{pdflscape}
\usepackage{multicol}  
\usepackage{multirow} 
\usepackage{booktabs}
\usepackage{array}
\usepackage{listings}  
\usepackage{xcolor} 
\usepackage{lipsum}       
\usepackage{changepage}   

\usepackage{diagbox}
\usepackage{stfloats}

\usepackage{algorithm}
\usepackage[noend]{algpseudocode}

\ifCLASSOPTIONcompsoc
  \usepackage[nocompress]{cite}
\else
  \usepackage{cite}
\fi

\usepackage{tikz}
\newlength\szg
\newcommand\quan[1]{%
	\settoheight\szg{#1}%
	\tikz[baseline]{\pgfmathparse{
			ifthenelse(#1 < 10, 1, ifthenelse(#1 < 100, 0.75, 0.5))
		}
		\let\hfs\pgfmathresult
		\node at (0,\szg/2) {\makebox[0em][c]{\scalebox{\hfs}[1]{#1}}};
		\draw (0,\szg/2) circle (\szg/2+0.35ex);
}}
\newcommand\hquan[1]{%
	\settoheight\szg{#1}%
	\tikz[baseline]{\pgfmathparse{
			ifthenelse(#1 < 10, 1, ifthenelse(#1 < 100, 0.75, 0.5))
		}
		\let\hfs\pgfmathresult
		\filldraw (0,\szg/2) circle (\szg/2+0.35ex);
		\node[white] at (0,\szg/2) {\makebox[0em][c]{\scalebox{\hfs}[1]{\textbf{#1}}}};
}}

\begin{document}
%
\title{Enabling Large-Reach TLBs for High-Throughput Processors by Exploiting Memory Subregion Contiguity}
%
%
%
%

\author{Chao Yu,
	Yuebin Bai,
	and Rui~Wang
	\IEEEcompsocitemizethanks{\IEEEcompsocthanksitem C. Yu, Y. Bai, and R. Wang are with the School of Computer Science, Beihang University, Beijing, China, 100191. 
		E-mail: \{yuchao, byb, wangrui\}@buaa.edu.cn
}}

\maketitle

\begin{abstract}

Accelerators, like GPUs, have become a trend to deliver future performance desire, and sharing the same virtual memory space between CPUs and GPUs is increasingly adopted to simplify programming. However, address translation, which is the key factor of virtual memory, is becoming the bottleneck of performance for GPUs. In GPUs, a single TLB miss can stall hundreds of threads due to the SIMT execute model, degrading performance dramatically. Through real system analysis, we observe that the OS shows an advanced contiguity (e.g., hundreds of contiguous pages), and more large memory regions with advanced contiguity tend to be allocated with the increase of working sets. Leveraging the observation, we propose \emph{MESC} to improve the translation efficiency for GPUs. The key idea of MESC is to divide each large page frame (2MB size) in virtual memory space into memory \emph{subregions} with fixed size (i.e., 64 4KB pages), and store the contiguity information of subregions and large page frames in L2PTEs. With MESC, address translations of up to 512 pages can be coalesced into single TLB entry, without the needs of changing memory allocation policy (i.e., demand paging) and the support of large pages. In the experimental results, MESC achieves 77.2\% performance improvement and 76.4\% reduction in dynamic translation energy for translation-sensitive workloads.

\end{abstract}




%
\IEEEpeerreviewmaketitle

\section{Introduction}
\label{sec:introduction}

Heterogeneous computing with accelerators like GPUs has become a trend to deliver desired performance in the Post-Moore's Law era \cite{kim2017heterogeneous,debenedictis2017s}.
 Traditional GPUs have seperate memory spaces. Before offloading computations to GPUs, all needed data should be copied from CPUs to GPUs through the Peripheral Component Interconnect express (PCI-e) bus, and the results will be copied back to CPUs after the computations are completed. The above process is handled explicitly by programmers. To avoid the data copying process and simplify programming, GPUs are integrated with CPUs on the same die and share the same virtual memory space with CPUs. In the industry, the Heterogeneous System Architecture (HSA) Foundation \cite{hsafoundation}, which was founded by companies including AMD, ARM, Imagination, Qualcomm, Samsung and Texas Instruments, relies on shared virtual memory to make heterogeneous accelerators a first-class computing devices.


Although the shared virtual memory can simplify programming and free GPUs from memory management, the virtual-to-physical address translation becomes an essential function of GPUs. The address translation, which is the key factor of virtual memory, has long been considered to be one of the main contributors to the performance overheads of CPUs as a TLB miss will cause multiple sequential memory accesses to walk the page table in x86-64 architecture, thus, a large number of works \cite{ganapathy1998general,seznec2004concurrent,pham2012colt,basu2013efficient,pham2014increasing,karakostas2015redundant,cox2017efficient,park2017hybrid} have concentrated on improving the efficiency of CPU address translation. Due to the differences in architecture and execution model, the address translation problem in GPUs is much more serious than in CPUs \cite{pichai2014architectural,power2014supporting,vesely2016observations,karnagel2017big,yoon2018filtering}. The state-of-the-art Memory Management Unit (MMU) designs \cite{power2014supporting,pichai2014architectural} optimized for GPUs are consisted of Translation Lookaside Buffers (TLBs), Page Table Walkers (PTWs) and Page Walk Cache (PWC). However, as the working sets and irregular memory accesses of applications are increased, the address translation is becoming the bottleneck of GPU performance due to the poor TLB reach \cite{vesely2016observations,karnagel2017big,ausavarungnirun2017mosaic,haria2018devirtualizing,yoon2018filtering,shin2018scheduling}. The address translation overhead due to TLB misses can cause slowdown of up to 13$\times$ for irregular GPU applications with divergent memory accesses \cite{karnagel2017big}, that is because a single TLB miss can stall hundreds of threads and the performance can be degraded significantly with the reduced thread level parallelism (TLP). Modern processors and operating systems (OSes) use 4KB base page coupling with 2MB or 1GB large pages to improve the reach of each TLB entry dramatically. To use large pages (e.g., THP \cite{arcangeli2010transparent} and libhugetlbfs \cite{mel2010libhugetlbfs}), the OS must guarantees all 4KB base pages in each allocated 2MB or 1GB page are contiguous in both physical and virtual spaces, thus, each TLB entry covers 2MB or 1GB contiguous memory, which increases the TLB hit ratio and reduces TLB miss latency significantly \cite{ganapathy1998general,seznec2004concurrent,arcangeli2010transparent,kwon2016coordinated}. 

Unfortunately, large pages are not the best choice to improve virtual memory efficiency. Large pages not only increase both I/O traffic and latency of page migration and demand paging, but also increase the probability of internal fragmentation.
Thus, many works \cite{pham2012colt,pham2014increasing,karakostas2015redundant,park2017hybrid,ausavarungnirun2017mosaic,haria2018devirtualizing} are proposed to disable large pages (see Section \ref{sec:largepage_shortcomings} for detailed discussion), and proposed their solutions based on base page to coalesce translations of contiguous base pages, which can improve TLB reach to various degrees. However, these solutions either need to change the default memory allocation policies (e.g., RMM \cite{karakostas2015redundant}, Mosaic \cite{ausavarungnirun2017mosaic} and DVM \cite{haria2018devirtualizing}), or can only coalesce limited pages (e.g., CoLT \cite{pham2012colt} and Cluster TLB \cite{pham2014increasing}).

\begin{figure}[hpt]
	\centering
	\includegraphics[width=3.3in]{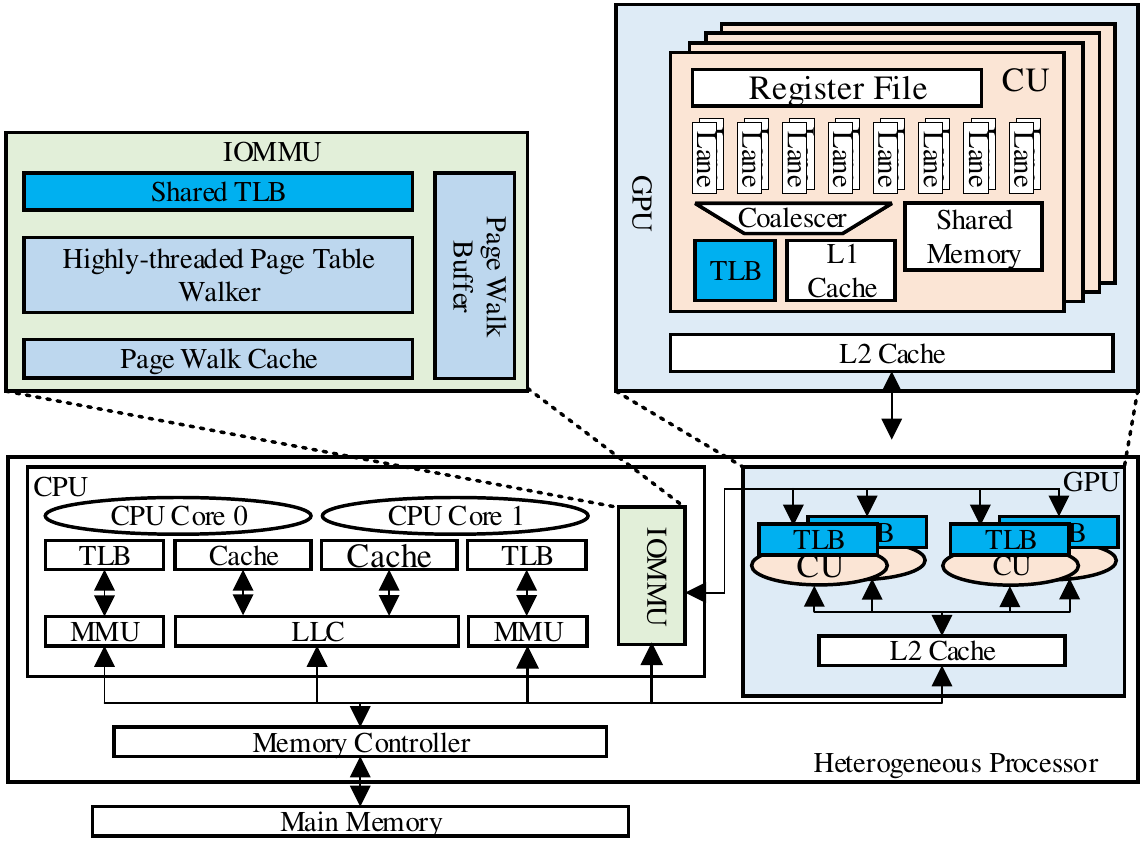}
	\caption{The baseline system architecture.}
	\label{fig:baseline_architecture}
\end{figure}

Nowadays, big-memory systems are becoming more and more popular and the working sets of applications are also growing. By analyzing the memory allocation of GPU applications in a real system, we have observed that the linux memory allocator shows an advanced contiguity \footnote{In comparison with the intermediate contiguity (e.g., tens of contiguous pages) exploited by CoLT \cite{pham2012colt}.} (e.g., hundreds of contiguous pages) and most of the working set can be covered by memory regions with advanced contiguity; besides, more memory regions with advanced contiguity tend to be allocated with the increase of working set. That means there exists an opportunity that the advanced contiguity can be exploited to improve the address translation efficiency without the support of large pages.

Leveraging the observation, we propose a new translation mechanism, called \emph{MESC}, exploiting the advanced contiguity to improve the address translation efficiency for high-throughput processors like GPUs. MESC divides each large page frame (2MB size) in virtual memory space into memory subregions with fixed size (i.e., 64 4KB pages), and stores contiguity information of both large page frames and their memory subregions in the unused bits in L2PTE. When walking the page table, these contiguity bits are analyzed to coalesce translations of pages in the corresponding contiguous subregion or large page frame. To improve the range of memory that can be coalesced, the contiguity between subregions in the same discontiguous large page frame is also checked. When the PTW is walking a L2PTE and analyses that the requested address resides in a discontiguous large page frame but belongs to a contiguous subregion, the head L1PTEs of all contiguous subregions around that subregion will be read to analysis the contiguity between these subregions. To reduce the memory accesses in checking the contiguity between subregions, we introduce a memory subregion cache (MSC) to store the contiguity information between memory subregions in the same large page frame. After the page table is walked, the coalesced translation of one or multiple subregions will be inserted into single subregion TLB entry. We also design an unified set-associative TLB structure based on the idea of way-partitioning to share the same TLB between regular TLB entries and subregion TLB entries, which avoids the need of a separate TLB for large pages.

To our knowledge, this paper is the first study to enlarge TLB reach by exploiting the advanced contiguity shown by OS without changing the default memory allocation policy and the support of large pages. In the experimental results, we observe that MESC achieves 77.2\% improvement in performance and 76.4\% reduction in dynamic translation energy for translation-sensitive workloads. By combining with CoLT \cite {pham2012colt}, which has been adopted by AMD, to also coalesce translations of several contiguous pages, MESC+CoLT can achieve 30\% reduction in dynamic translation energy for translation-insensitive workloads, although the performance is not improved obviously.
\section{Background and Motivation}
\label{sec:background}

\subsection{GPU Architecture and Address Translation}

Figure \ref{fig:baseline_architecture} depicts the baseline heterogeneous architecture, which is similar with \cite{power2014supporting} and \cite{yoon2018filtering}, we used in this paper. GPU and CPU share the same physical memory and virtual memory space, and cache coherence is maintained between CPU and GPU. A GPU usually contains several to dozens of Computing Units (CU) in AMD terminology \footnote{The CU is also called Execution Unit (EU) in Intel terminology and Streaming Multiprocessor (SM) in Nvidia terminology. We use the AMD terminology in this paper.}. A CU has a set of lanes (SIMD units), which are function units including ALU, SFU, etc. Each lane can finish one instruction from a \emph{workitem} or \emph{thread} in a cycle if no stall occurs. In GPUs, instruction stalls and long latency memory accesses are hidden by fast switching among massive concurrent threads, thus, each CU has a large register file to hold contexts of all active threads. GPUs are also equipped with a shared memory for each CU to support inter-thread communication within a group of threads called \emph{workgroup} or \emph{thread block}. On execution, threads in each workgroup are grouped into smaller groups called \emph{wavefront} or \emph{warp}, which is the basic unit of scheduling, and each wavefront consists of 32 or 64 threads. All threads in the same wavefront are executed in a lockstep fashion, thus, the stall of a thread will leads to all threads in the same wavefront are stalled. To reduce the latency of memory accesses, each CU has a private L1 cache and all CUs share a L2 cache, and memory requests, which reside in the same cache line, from the same wavefront instruction are coalesced to a single access to improve memory access efficiency. 

As CPU and GPU share the same virtual memory space in heterogeneous computing architecture like HSA \cite{hsafoundation}, GPU virtual addresses (VAs) should be translated to physical addresses (PAs) by walking the same x86-64 page table as the CPU before data can be actually accessed from cache or main memory. The IO Memory Management Unit (IOMMU) \cite{vesely2016observations} in the processor is responsible for walking the page table for address translation requests of accelerators including the GPU. Although accelerators are connected to IOMMU through the interconnection network, the communication protocol between IOMMU and accelerators is PCIe \cite{vesely2016observations}. Since each page table walk can incur up to four memory accesses in x86-64 architecture, which can degrade GPU performance significantly, the GPU has its own L1 per-CU TLBs to cache recently used address translations to avoid page table walk on every address translation request. On a per-CU TLB miss, the translation request is sent to IOMMU. The IOMMU consists of a TLB, highly-threaded page table walkers (PTWs), page walk buffer (PWB) and page walk cache (PWC). The IOMMU TLB, which can be considered as a L2 shared GPU TLB, is shared across all of the CUs and is looked up upon a request's arrival. If the request is missed in the IOMMU TLB, a free PTW is chose to walk the x86-64 page table in main memory. To service bursts of TLB misses \cite{power2014supporting}, the PTW is usually highly-threaded (e.g. 8-32 threads) to support concurrent page table walks \cite{power2014supporting,yoon2018filtering,shin2018scheduling}. If there is no free PTW currently, the arrived request is put in the PWB to wait for a free PTW. Once a page table walk request is finished, the IOMMU sends the translation result back to GPU, and the corresponding PTW will be released and then assigned to a request in PWB. To reduce page walk latency, the PWC is used in IOMMU to cache the top three levels of the x86-64 page table, which can reduce the number of memory accesses for a page table walk to one.

\subsection{The Need of Large TLB Reach for GPUs}

\begin{figure}
	\centering
	\includegraphics[width=3.3in]{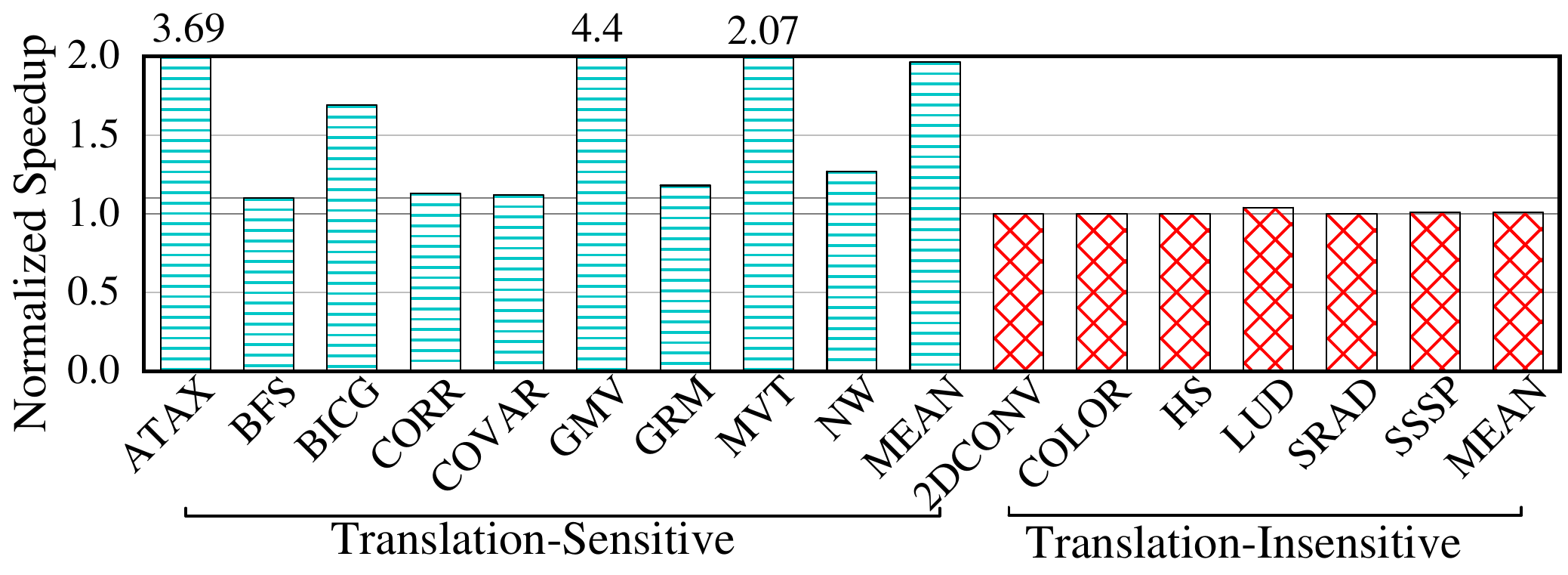}
	\caption{Speedup of GPU performance with THP.}
	\label{fig:bl_translation}
\end{figure}

Address translation overheads can severely degrade the performance of GPU applications especially irregular applications \cite{vesely2016observations,karnagel2017big,ausavarungnirun2017mosaic,haria2018devirtualizing,yoon2018filtering,shin2018scheduling}. To quantify the effect of address translation, we compare the performance of GPU applications using a baseline MMU design and a THP design with a full-system heterogeneous simulator gem5-gpu \cite{power2015gem5}. The baseline MMU design is based on the work from \cite{power2014supporting} and the THP design is the scenario using 2MB transparent huge page \cite{arcangeli2010transparent}. All evaluated applications are selected from Polybench \cite{pouchet2012polybench}, Rodinia \cite{che2009rodinia} and Pannotia \cite{che2013pannotia}. We classify all selected applications into 2 categories: \emph{translation-sensitive} and \emph{translation-insensitive}. If an application can obtain at least 10\% of the speedup with the THP design, it is a translation-sensitive application, otherwise, translation-insensitive application. Figure \ref{fig:bl_translation} shows the normalized speedup of GPU performance using THP design to baseline MMU design. With less address translation overheads, the GPU performance can be accelerated up to 4.4$\times$ and the average GPU performance speedup is 1.96$\times$ for translation-sensitive applications. Because the address translation overhead is insignificant for translation-insensitive applications, the GPU performance speedup is negligible for the selected translation-insensitive applications. The above results imply that address translation can indeed reduce GPU performance seriously and an efficient method should be used to improve address translation efficiency.

\begin{figure}
	\centering
	\includegraphics[width=3.3in]{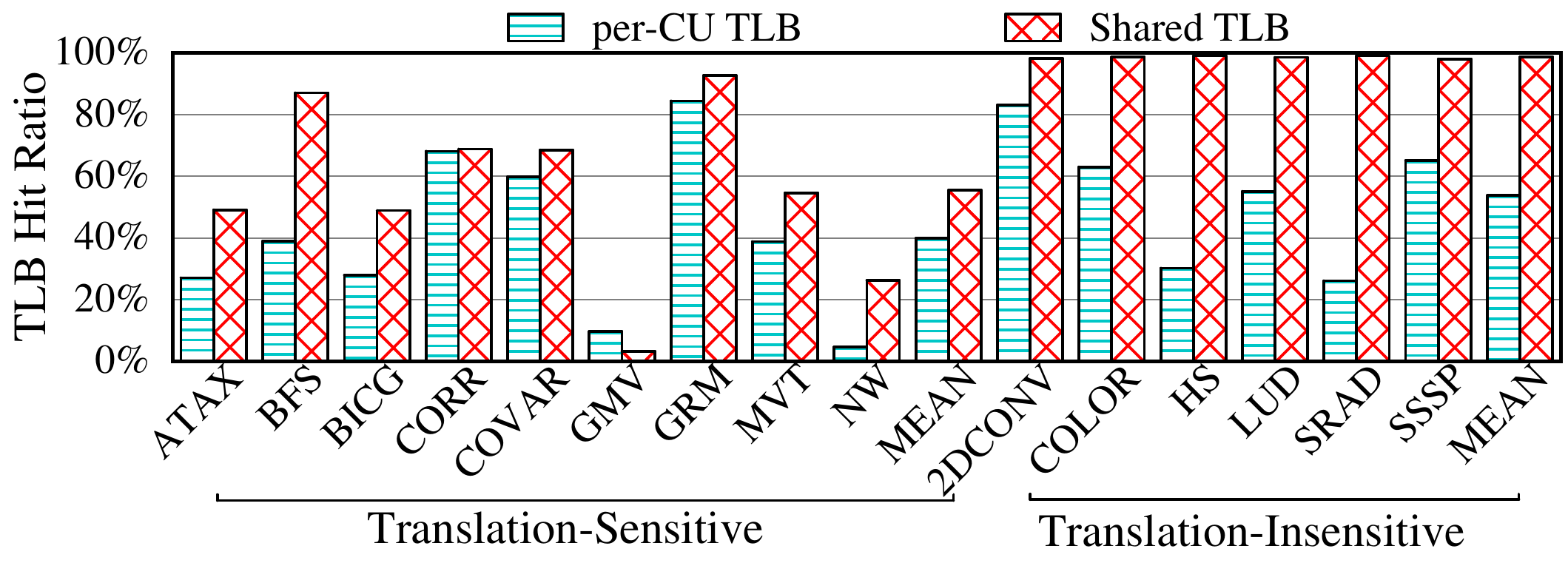}
	\caption{Hit ratios of per-CU TLB and shared IOMMU TLB.}
	\label{fig:bl_tlbhitratio}
\end{figure}


To further analysis the reason of performance degradation due to address translation, we record the TLB hit ratio and memory bandwidth consumption for each application. Figure \ref{fig:bl_tlbhitratio} shows the hit ratios of both per-CU TLB and shared IOMMU TLB for each application. We can find that hit ratios of both per-CU TLB and shared TLB are generally very low for translation-sensitive applications, and the average hit ratios are 39.91\% and 55.42\% respectively. As a contrast, although the average TLB hit ratio of per-CU TLB for translation-insensitive applications is only 53.75\%, the average TLB hit ratio of shared TLB reaches 98.55\%, which can avoid most page table walk requests. The low TLB hit ratio is directly reflected in the competition for memory bandwidth. Due to the low TLB hit ratio of translation-sensitive applications, the page table walk bandwidth of translation-sensitive applications accounts for a fairly high percentage, 26.70\% on average in our evaluation, of the total GPU memory bandwidth consumption, which may cause intense competition for memory bandwidth and degrade GPU performance seriously. Since address translation is in the critical path of pipeline and TLB miss penalty is high, it is critical to improve address translation efficiency. Currently, the widely adopted approach to improve address translation efficiency is improving TLB reach.

\subsection{Large Pages are Not Silver Bullet}
\label{sec:largepage_shortcomings}
The most direct and simple method to improve TLB reach is coupling 4KB base page with large pages (e.g. 2MB and 1GB page sizes), which are already supported by current architectures (e.g. x86-64) and OSes (e.g. Linux). However, there are several shortcomings impeding the applicability of large pages. 

First, the NUMA and emerging hybrid/heterogeneous memory (HM) architectures prefer to put frequently accessed pages to fast near memory \cite{jevdjic2013stacked,dashti2013traffic,agarwal2017thermostat,yu2017banshee}. If the page granularity is large, the frequent memory migrations between fast and slow memories need more I/O traffic and may exhaust the memory bandwidth, and that is detrimental to performance. 
Second, demand paging is one of the main features of virtual memory, which enables applications to execute without all pages currently allocated or resident in main memory. When the requested pages are currently not allocated or resident in main memory, page faults will be triggered. Then the unallocated pages will be allocated and the nonresident pages will be transferred from secondary storage to main memory through the system I/O bus. With large pages, page faults incur much larger overhead than base page and GPU performance can be degraded seriously \cite{vesely2016observations,zheng2016towards,ausavarungnirun2017mosaic}. Third, as memory is allocated in page granularity, larger pages exacerbate the problem of internal fragmentation and leads to the unavailability of large pages. In such case, frequent memory compaction \cite{jonathan2010memory} will be performed by OS to generate more large pages, which has been observed that it has become a source of poor performance especially in long-running systems \cite{jonathan2010memory,jonathan2017proactive,panwar2018making}, and many common scenarios, such as big data platforms (e.g., Hadoop \cite{disablethphadoop}) and database services (e.g., Oracle \cite{disablethporacle} and MongoDB \cite{disablethpmongodb}), recommend to disable THP. Forth, supporting multiple page sizes will increase the overheads of both software and hardware. In the OS, special code should be maintained to manage pages with different sizes, and large pages must be aligned in both virtual and physical memory spaces. Besides, split TLBs \cite{papadopoulou2015prediction,cox2017efficient}, which can lead to underutilized TLBs \cite{cox2017efficient}, are needed to cache translation information for different page sizes. 

Based on the above challenges, many works \cite{talluri1994surpassing,pham2012colt,basu2013efficient,pham2014increasing,karakostas2015redundant,park2017hybrid,ausavarungnirun2017mosaic,haria2018devirtualizing} are proposed to disable large pages and exploit the contiguity of base pages to improve the address translation efficiency. However, we found there still exist some restrictions or disadvantages when applied to GPUs. Some solutions need to change the default memory allocator to allocate very large contiguous memory regions (e.g., RMM \cite{karakostas2015redundant} and DVM \cite{haria2018devirtualizing}) or implement a completely different memory management strategy (e.g., Mosaic \cite{ausavarungnirun2017mosaic}); some solutions have high TLB synchronization overhead (e.g., Hybrid TLB \cite{park2017hybrid}); some solutions lack flexibility (e.g., Direct Segments \cite{basu2013efficient}); other solutions can only coalesce limited pages (e.g., CoLT \cite{pham2012colt}, Sub-block TLB \cite{talluri1994surpassing} and Cluster TLB \cite{pham2014increasing}), lacking scalability. Thus, we think a new solution is still needed, especially for GPUs, to improve the address translation efficiency while avoiding above problems.

\section{Observation and Opportunity}
\label{sec:observation}


%

\begin{figure}
	\centering
	\includegraphics[width=3.3in]{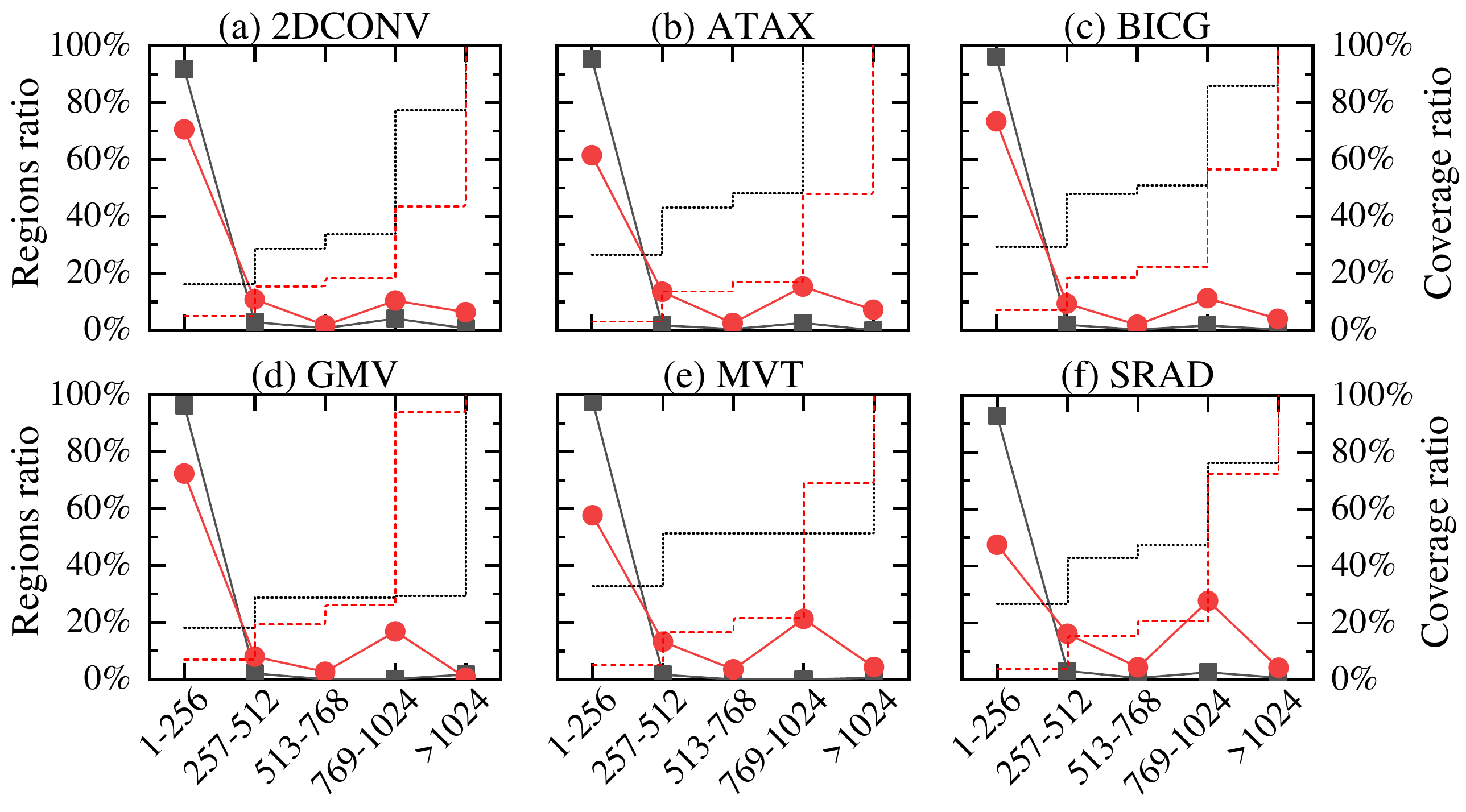}
	\caption{Distributions of contiguous memory regions. Solid lines are distributions of region sizes; dashed lines are cumulative distributions of the memory footprint covered by regions with various contiguous pages. Black lines represent small working sets and red lines represent larger working sets.}
	\label{fig:apps_num_cov}
\end{figure}

To observe the contiguity shown by the default memory allocator and analysis the distribution of contiguous memory regions, we have made a set of experiments. We ran multiple GPU applications on a long-running system, which is equipped with 32GB DRAM, to pressure the system memory, and collect their virtual to physical memory address mapping information using Linux's \emph{pagemap} interface. As the memory data used by GPU is usually allocated in heap, only memory segments located in heap are considered. THP is disabled during the experiments.

Figure \ref{fig:apps_num_cov} shows the distribution of contiguous memory regions, whose virtual addresses (VAs) are contiguously mapped to physical addresses (PAs), with various sizes \footnote{Page is the smallest contiguous memory region.} under small and larger working sets. In Figure \ref{fig:apps_num_cov}, the x-axis represents the scopes of memory regions (e.g., "1-256" represents contiguous memory regions composed of 1 to 256 base pages), and the left y-axis and right y-axis represent region number ratio and cumulative coverage ratio respectively. From the results shown in Figure \ref{fig:apps_num_cov}, we can see that the memory allocator tends to allocate more small memory regions under small working set (e.g., the ratios of memory regions within "1-256" are up to 90\% in most cases), and the ratio of larger memory regions is improved (e.g., the ratio of memory regions within "769-1024" is up to 27.72\% in SRAD) under larger working set. However, higher ratio of small memory regions does not mean that most of the memory is covered by these small memory regions. On the contrary, we can find that the less large memory regions usually cover most of the memory footprint under both small and larger working sets, which means that the default memory allocator shows an advanced contiguity (e.g., hundreds of contiguous pages). From above analysis, we can conclude the following observation.

\textbf{Observation:} \emph{The default memory allocator shows an advanced contiguity (e.g., hundreds of contiguous pages) and most of the memory footprint can be covered by larger contiguous memory regions; besides, more large contiguous memory regions tend to be allocated with the increase of working sets.}


The above observation implies that there exists an opportunity that the existing advanced contiguity generated by the OS can be exploited to improve the address translation efficiency. To exploit the advanced contiguity, the challenge that \emph{how to manage the contiguity information of memory regions with various sizes} should be addressed. The various contiguous memory regions can be any size such as 64 base pages, 320 base pages, 512 base pages, 739 page pages and so on. Previous work \cite{karakostas2015redundant} directly stores the base addresses and ranges of all contiguous regions with a range table in memory as page table and caches contiguity information in an extra fully associative range TLB. But it requires the OS to allocate a small number of very large contiguous memory regions. In this paper, we divide the virtual memory space into \emph{subregions} with fixed sizes (i.e., 64 base pages). Thus, each large page frame (2MB) in virtual memory space is composed of 8 subregions, and the contiguity of large memory regions with various sizes can be exploited using subregion as unit. One of the main advantages of our approach is that page contiguity can be exploited in a smaller granularity than large page. Thus, there is no demand that the memory allocator should allocate large pages 2MB aligned, since the contiguity of memory regions within a discontiguous large page can also be exploited. 

Although both our subregion based method and CoLT \cite{pham2012colt} exploit the existing contiguity generated by OS memory allocator, the main difference between CoLT and our method is the number of contiguous virtual-to-physical address translations that can be coalesced. As PTW reads PTEs in units of cache lines, CoLT mainly tries to coalesce translations from L1PTEs located in the same cache line. For example, assume that the cache line size is 128B, thus each cache line contains 16 L1PTEs and at most 16 contiguous translations can be coalesced into single TLB entry. In practical, less translations (e.g., 4 or 8 contiguous translations) are coalesced to balance TLB reach and conflict misses. Different from CoLT, our method exploits the advanced contiguity generated by OS memory allocator, which is a completely different method from CoLT. By dividing each large page frame into memory subregions with fixed size, our method can coalesce address translations of up to 512 pages into single TLB entry.


\section{MESC Design and Implementation}
\label{sec:pasc_design}
\subsection{Memory Subregion Coalescing}

Although large pages are usually disabled due to shortcomings discussed in Section \ref{sec:largepage_shortcomings}, the default memory allocator (e.g., buddy allocator) in OS still shows forms of contiguity. Previous work CoLT \cite{pham2012colt} use PTE coalescing to exploit the intermediate contiguity (e.g., tens of contiguous pages), which can only cover limited number of pages in each TLB entry. We also have found that the default memory allocator also shows a form of advanced contiguity (e.g., hundreds of contiguous pages), which can cover most of the memory footprint, and this feature becomes more apparent with the increase of working sets. Thus the advanced contiguity can be exploited in a coarse-grained method to coalesce more pages.

In this paper, we divide each large page frame in virtual memory space into \emph{subregions} with fixed size. The contiguity of each memory region can be represented by the contiguity of subregions in it. After memory is allocated, the page table will be scanned and the contiguity information of subregions will be stored back to the page table. When the page table being walked, the contiguity information will be extracted to judge whether the requested address belongs to a contiguous region. If so, the coalesced translation information will be cached in TLB.

\begin{figure}
	\centering
	\includegraphics[width=3.3in]{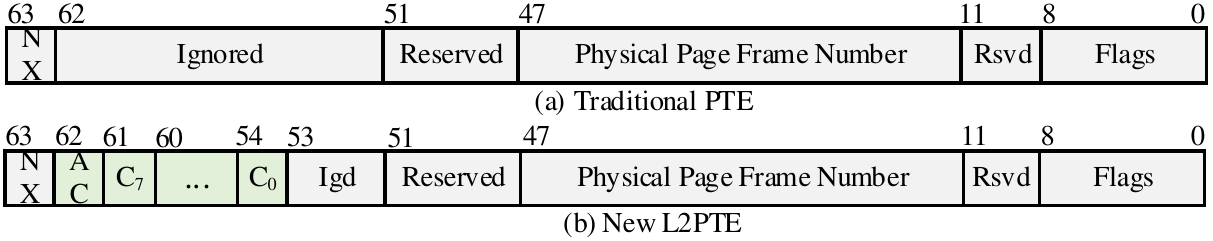}
	\caption{The translation PTE format and modified L2PTE format.}
	\label{fig:pte_format}
\end{figure}

As there are some unused bits in the PTE, we store the contiguity information in those unused bits. Figure \ref{fig:pte_format} shows the modified L2PTE format we used to store the contiguity information. For a subregion, if all VAs in it are contiguously mapped to PAs, the corresponding $C_{x}$ bit, which is initially set to 0, in its L2PTE is set to 1 indicating that it is contiguous (i.e., contiguously mapped). For example, if the second subregion of a large page frame is contiguous, the $C_{1}$ bit in its L2PTE is set to 1. Because even if all the $C_{x}$ bits of a L2PTE are set to 1, it does not mean the corresponding large page frame is also contiguously mapped, so we add an $AC$ bit, which is initially set to 0, in each L2PTE. If a large page frame is contiguous, the corresponding $AC$ bit will be set to 1. Besides, the contiguity between subregions in each discontiguous large page frame is also explored (see Section \ref{sec:page_table_walking}) to coalesce as many translations as possible. It should be noted that although memory subregions and large page frames are aligned in virtual memory space, the physical memory space has no such restriction. For example, a virtual large page frame from 0x80000 to 0x801FF can be contiguously mapped to a physical memory region from 0x6000A to 0x60209, which is not 2MB aligned. While our memory subregion based method can exploit the advanced contiguity in subregion granularity, the intermediate contiguity in discontiguous memory subregions can also be exploited by combining with CoLT (see Section \ref{sec:combine_colt}).


\subsection{Page Table Walking}
\label{sec:page_table_walking}

\begin{figure}
	\centering
	\includegraphics[width=3.3in]{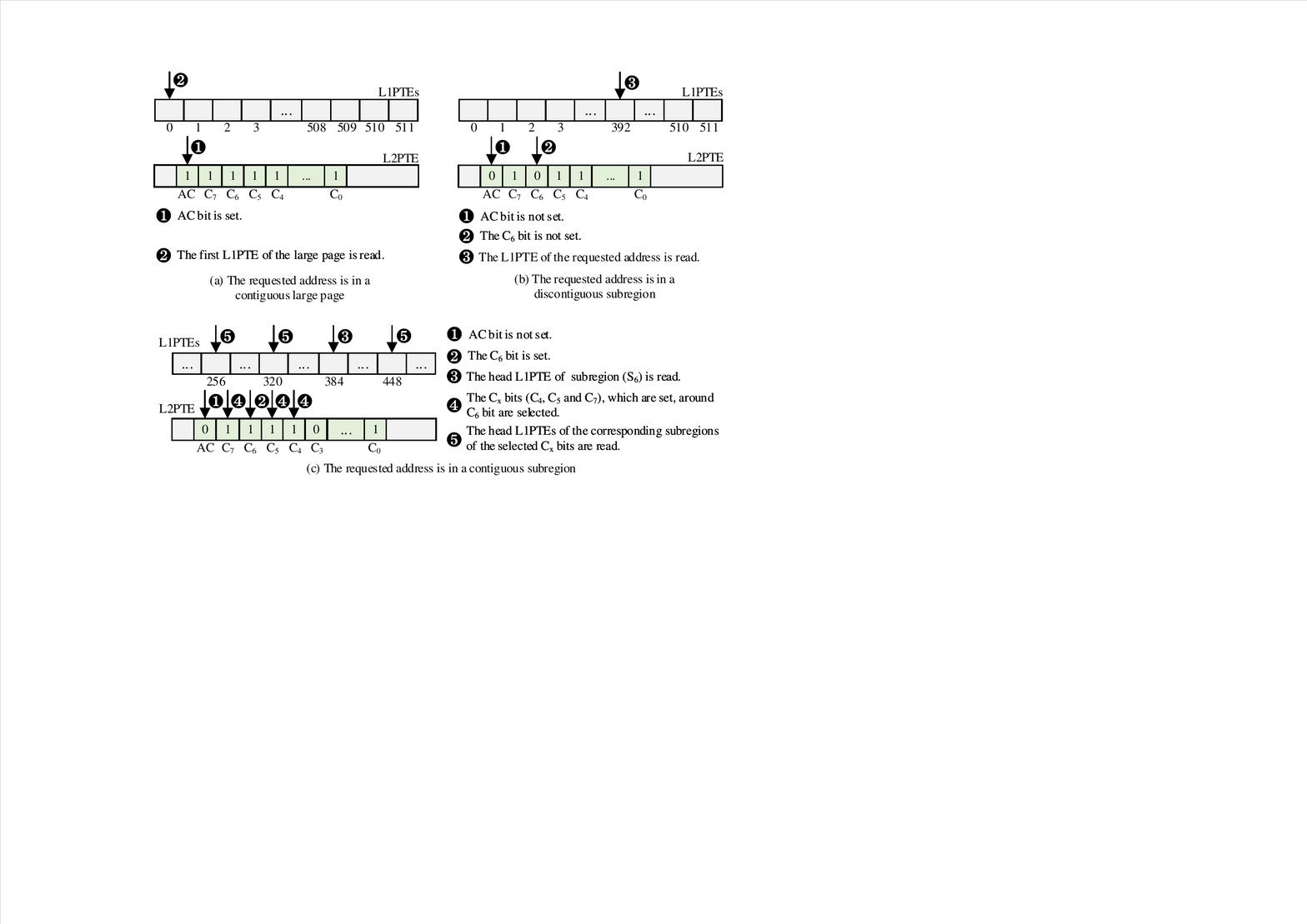}
	\caption{Three different modes of page table walking with different flags (L3PTE and L4PTE are not shown). The VFN of the requested VA in the three cases is assumed to be 0x80188.}
	\label{fig:ptw_cases}
\end{figure}

On a per-CU TLB miss, the translation request is sent to IOMMU. If the IOMMU TLB is also missed, the page table will be walked by the PTW to locate the PA mapped to the corresponding VA. There are three different modes of page table walking shown in Figure \ref{fig:ptw_cases}. When walking the L2PTE, the $AC$ bit of L2PTE will first be checked to see if the requested address belongs to a contiguous large page frame. If $AC$ is set (see Figure \ref{fig:ptw_cases}(a)), which means the corresponding large page frame is contiguously mapped, the PTW reads the first L1PTE pointed by the L2PTE to get the start physical page frame number (PFN) of mapped memory region. Since the large page frame is contiguously mapped, the mapped PFN of the requested address can be obtained by adding the offset between the start base virtual page frame number (VFN) of the large page frame represented by the L2PTE and the VFN of the requested address to the start PFN. Then the translation is completed and the translation result is returned to GPU lanes, and the coalesced translation is inserted to the TLB. If $AC$ is not set, then the corresponding $C_{x}$ bit will be checked. If the corresponding $C_{x}$ bit is not set (see Figure \ref{fig:ptw_cases}(b)), the corresponding L1PTE of the requested address is read. Then the translation result is returned to GPU lanes and inserted to the TLB. If the corresponding $C_{x}$ bit is set (see Figure \ref{fig:ptw_cases}(c)), which means the requested address resides in a contiguous subregion, the first (head) L1PTE of the subregion is read; besides, the contiguity between that subregion and other subregions in the same large page frame is also checked. As address translation is in the critical path of execution, the generated translation result is returned to GPU lanes immediately after the head L1PTE of the current subregion is read. Then, the head L1PTEs of all contiguous subregions around the subregion are read, and the coalesced address translation of these consecutive subregions is inserted to the TLB. How address translations of multiple contiguous subregions are coalesced into one TLB entry is discussed in below section. 

According to above analysis, checking the contiguity between subregions is only happened when the large page frame contains at least one discontiguous subregion and the requested address resides in a contiguous subregion. As the head L1PTE of the subregion containing the requested address has first been read, the head L1PTEs of up to 6 other contiguous subregions in the same large page frame also should be read, the overhead of which can be expensive. In this paper, we introduce a set-associative memory subregion cache (MSC) to IOMMU to filter the memory accesses when checking the contiguity between subregions in the same large page frame.

\begin{figure}
	\centering
	\includegraphics[width=3.3in]{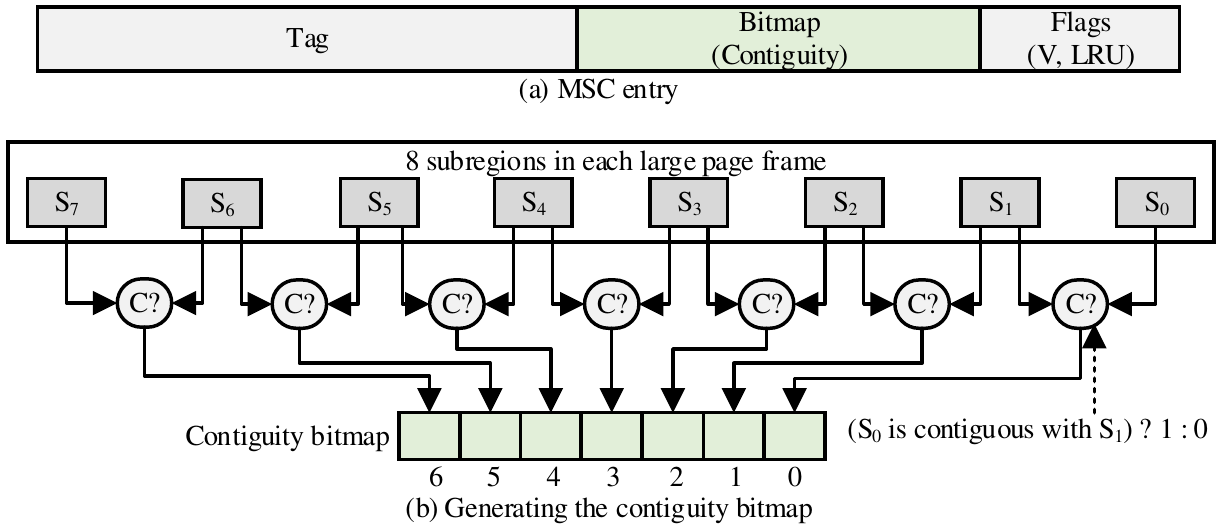}
	\caption{The contiguity information between memory subregions in a large page is stored as a bitmap in memory subregion cache.}
	\label{fig:memory_subregion_cache}
\end{figure}

The entry of MSC and how the contiguity information between memory subregions is generated are shown in Figure \ref{fig:memory_subregion_cache}. The contiguity information between the 8 subregions in each large page frame is stored as a 7-bit bitmap and each entry of MSC stores the contiguity information of one large page frame. In MSC, if the contiguity exists in the interior of subregions $S_{i}$ ($0 \le i \le 7$) and $S_{i+1}$ as well as between them, the $i$th bit will be set to 1. For example, if both the $C_{i}$ bit and $C_{i+1}$ bit in L2PTE are 1, and the PFN in the head L1PTE of subregion $S_{i}$ is 64 less than that of subregion $S_{i+1}$, then, the $i$th bit of the bitmap will be set to 1 when the contiguity information is inserted to MSC. When walking page table and the contiguity between subregions in the same large page frame should be checked (see Figure \ref{fig:ptw_cases}(c)), the MSC lookup is performed. If there is a matching entry in MSC, the contiguity information is extracted from the contiguity bitmap of that entry and then the coalesced address translation is inserted to the TLB. If there is no matching entry in MSC, the head L1PTEs of all other contiguous subregions in the same large page frame are read and the generated contiguity information of the large page frame is inserted to MSC.


\begin{figure*}
	\centering
	\includegraphics[width=7in]{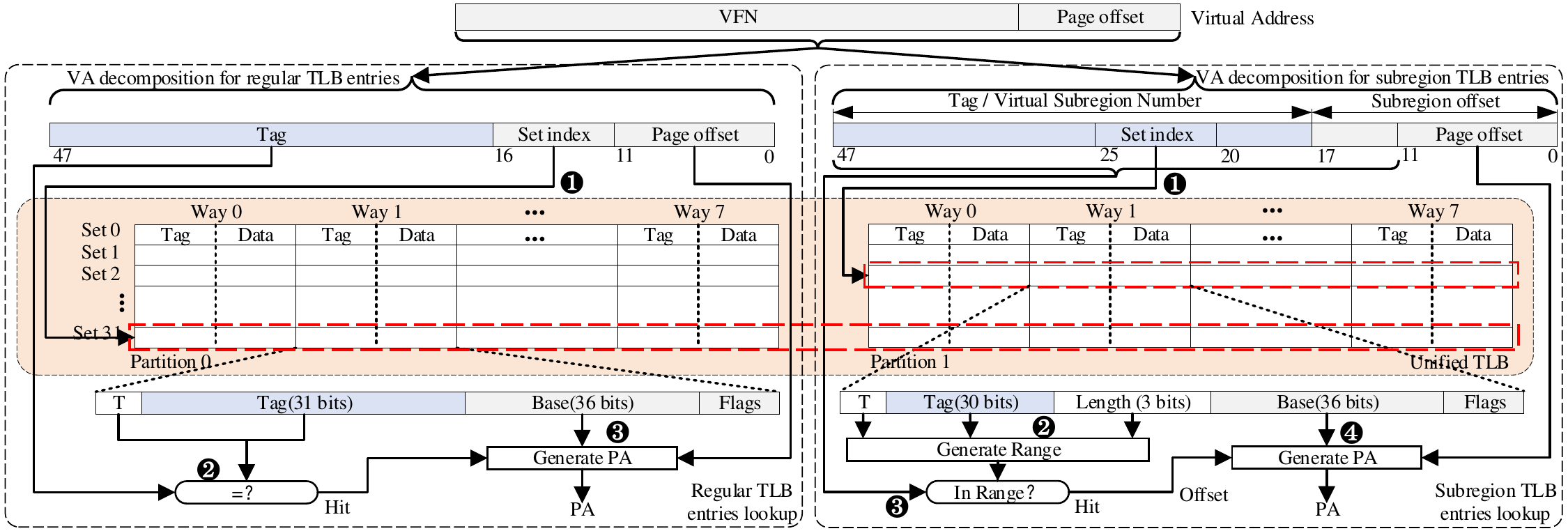}
	\caption{The unified set-associative TLB structure (32-set) of MESC. The top and bottom of the picture show VA decompositions and TLB lookups, respectively, for different types of TLB entries.}
	\label{fig:unified_tlb}
\end{figure*}

\subsection{TLB Lookups}
\label{sec:tlb_lookups}

To store coalesced address translations, the TLB structure should be modified. As per-CU TLBs are integrated into each CU and GPU performance is sensitive to per-CU TLBs' access latency, the per-CU TLBs must keep simple and fast. The IOMMU TLB is large and already far from CUs and can tolerate higher access latency, therefore, the support for MESC is only added to the IOMMU TLB to coalesce more translations. 

\textbf{TLB Entries.} 
The modified TLB entries for regular pages and coalesced subregions are shown in the bottom of Figure \ref{fig:unified_tlb}. Both the two types of entries require an additional bit (field $T$) to record the entry type, and the $T$ values of regular TLB entry and subregion TLB entry are 0 and 1 respectively. Besides, subregion TLB entry uses a 3-bit length field to record the contiguity of up to 8 consecutive subregions in the same large page frame. As the set index bits for subregion TLB entry are VA[25-21] instead of VA[22-18] out of the subregion offset bits (explained below), we take the whole virtual subregion number (VSN) (30 bits) as tag \footnote{In implementation, the set indexing bits can further be eliminated from TLB tag.} for the simplicity of description. Although subregions in VA space are required to be 64-page aligned, only 4KB page alignment in PA space is required for the mapped contiguous physical memory region of each contiguous subregion. Thus, the data field of subregion TLB entry still needs to store the mapped base PFN (36 bits) of memory region covered by the TLB entry. Some other bits are also needed in both types of entries for validation and permissions.

\textbf{Unified Set-associative TLB.} 
As there exist two types of TLB entries simultaneously, we design an unified set-associative TLB structure (although split TLBs are also applicable) shown in Figure \ref{fig:unified_tlb} for MESC to accommodate the two types of entries in the same TLB. The unified TLB structure is based on the classic idea of way-partitioning. In Figure \ref{fig:unified_tlb}, the 16-way TLB is divided into two partitions. Because a subregion TLB entry can cover much larger memory region (at least 64 base pages and up to 512 base pages) than a base page TLB entry, subregions need a smaller TLB than base pages. Thus, in our unified TLB, base pages can use all the ways in the two partitions while subregions can only use the ways in a designated partition. The way-partitioning based unified TLB eliminates the need of assigning a separate TLB for subregions, saving chip area overhead.

\textbf{VA Decomposition.} 
Since different TLB entries are resident in the same TLB and it is unknown whether a requested VA is in a contiguous subregion or not, both the two types of entries may be searched. The VA decompositions for the two types of TLB entries are shown in the top of Figure \ref{fig:unified_tlb}. The VA decomposition for regular TLB entries is the same as traditional TLB, and VFN[4-0] (bits 16 to 12) are used for set selection and other bits of VFN are used for tag comparison. For subregion TLB entries, the set selection bits are different. As each subregion TLB entry can cover 1 to 8 consecutive subregions, traditional method choosing bits 4 to 0 of the VSN for set selection would map translations of consecutive subregions to consecutive sets, which prevents subregions coalescing. To enable the address translations of up to 8 consecutive subregions in the same large page frame to be coalesced into a single entry and ensure subregions of consecutive large page frames are mapped to different sets of the TLB, the set selection bits are left-shifted by 3 bits to VSN[7-3] (bits 25 to 21).

\textbf{Lookup Operation.} 
On a per-CU TLB miss, the shared TLB is looked up with the process illustrated in Figure \ref{fig:unified_tlb}. 
As subregion TLB entries cover much larger footprint than regular TLB entries, the subregion TLB partition is first looked up to check if the translation request belongs to any contiguous subregion. Different from regular TLB entries, subregion TLB entries are only resident in a designated partition (partition 1 in Figure \ref{fig:unified_tlb}). When looking up subregion TLB entries, the desired set of the designated partition is selected with the left-shifted indexing bits \hquan1. Then the address range of each subregion TLB entry (\emph{T} = 1) is calculated \hquan2. According to above analysis, although the tag field in each subregion TLB entry is VSN, the memory coverage is calculated in VFN.
The lower bound VFN (i.e., base VFN) $\mathit{VFN_{lower}}$ and upper bound VFN $\mathit{VFN_{upper}}$ of the memory coverage are calculated using Equations \ref{eq:cal_vfn_lower} and \ref{eq:cal_vfn_upper} respectively:
\begin{equation}
\mathit{VFN_{lower}}=Tag << 6,
\label{eq:cal_vfn_lower}
\end{equation}
\begin{equation}
\mathit{VFN_{upper}}=((Tag+Length)<<6) | 0x3F.
\label{eq:cal_vfn_upper}
\end{equation}
If the VFN of the requested address is located in the coverage of any subregion TLB entry \hquan3, the translation hits in TLB. Then the offset of the requested VFN to the base VFN of the memory region covered by the entry, the data field of the entry and the page offset of the requested VA are used to generate PA \hquan4. 
If the subregion TLB entries miss, the regular TLB entries are then looked up. The process of looking up regular TLB entries is almost the same as traditional TLB. 
If both the subregion TLB entries and regular TLB entries miss, a page table walk request will be generated (see Section \ref{sec:page_table_walking}). 

\begin{figure}
	\centering
	\includegraphics[width=3.3in]{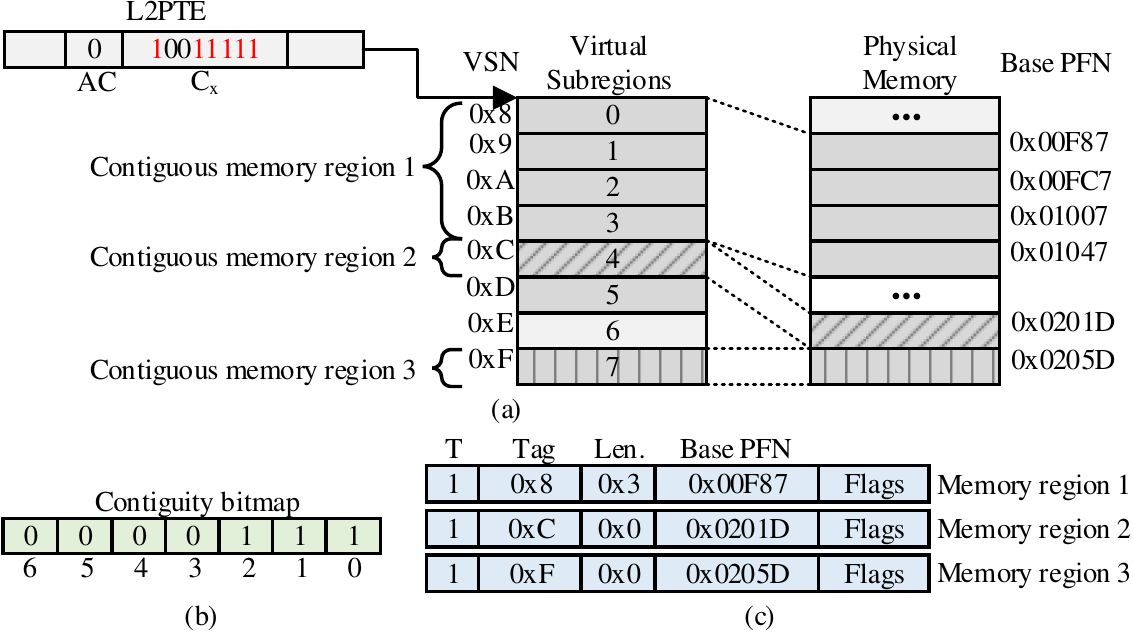}
	\caption{Memory subregion based TLB coalescing example. (a) Three contiguous memory regions. (b) Generated contiguity information in MSC. (c) Coalesced TLB contents of the three memory regions.}
	\label{fig:tlb_coalescing_example}
\end{figure}

\textbf{TLB Coalescing Example.}
Now we take an example shown in Figure \ref{fig:tlb_coalescing_example} to illustrate how address translations of consecutive memory subregions are coalesced in single subregion TLB entry. Figure \ref{fig:tlb_coalescing_example}(a) shows all $C_{x}$ bits except bits $C_{5}$ and $C_{6}$ in L2PTE are 1, which means there are 6 contiguous subregions in the virtual large page frame represented by the L2PTE and the large page frame seems to have 2 subregion based contiguous memory regions. Further, by analysing the base PFNs of each contiguous subregions, although subregions $S_{0}$ to $S_{4}$ are all contiguous, there is no contiguity between subregions $S_{3}$ and $S_{4}$. Thus, the large page frame actually has 3 subregion based contiguous memory regions. When the page table is walked, the contiguity information shown in Figure \ref{fig:tlb_coalescing_example}(b) between each subregions in the large page frame is generated and inserted to MSC. In addition, the coalesced address translation of the contiguous memory region corresponding to the requested address is inserted to the TLB. The coalesced subregion TLB entries of the three contiguous memory regions are shown in Figure \ref{fig:tlb_coalescing_example}(c). In these coalesced TLB entries, the tag values are the base VSNs of each contiguous memory region. As the lengths of each contiguous memory region are 4, 1 and 1 respectively, the length fields of these coalesced TLB entries are 0x3, 0x0, and 0x0 respectively. The data fields are the mapped base PFNs of these contiguous memory regions, which are 0x00F87, 0x0201D and 0x0205D respectively. 

\subsection{OS Support}
\label{sec:ossupport}
MESC doesn't rely on the OS to allocate large contiguous physical memory regions (e.g. eager allocation \cite{karakostas2015redundant,haria2018devirtualizing}). The main purpose of MESC is to exploit the existing advanced contiguity of the default memory allocator. Thus, only minor modifications to the OS are needed to scan page table of each process and store the subregion based contiguity information in page table.

\begin{algorithm}[h]
	\caption{Page table scanning algorithm}
	\label{alg:pagetable_scanning}
\begin{algorithmic}[1]
	\For{each $L2PTE$ in page table} 
		\State $allsubcont$ $\leftarrow$ true
		\State $subregionlist$ $\leftarrow$ all subregions in the large frame
		\State $presubregion$ $\leftarrow$ $subregionlist[0]$
		\For {each $cursubregion$ in $subregionlist$}
			\State $cursubcont$ $\leftarrow$ true
			\State $//$ \textit{Exploit the contiguity of current subregion}
			\If{not SubCont($cursubregion$)}
				\State $cursubcont$ $\leftarrow$ false
			\Else
				\State SetCxBit($L2PTE$,$cursubregion$,1)
			\EndIf
			
			\State $//$ \textit{Exploit the contiguity between subregions}
			\State $intersubcont$ $\leftarrow$ InterSubCont($cursubregion$, $presubregion$)
			\If{not $cursubcont$ \textbf{or} not $intersubcont$}
				\State $allsubcont$ $\leftarrow$ false
			\EndIf
			\State $presubregion$ $\leftarrow$ $cursubregion$
		\EndFor
		\If{$allsubcont$}
			\State SetACBit($L2PTE$,1)
		\EndIf
	\EndFor
\end{algorithmic}
\end{algorithm}

\textbf{Identifying Subregion Contiguity.}
To identify the contiguity of subregions, the page table should be scanned and the information of identified contiguous subregions will be stored in page table. The process of identifying subregion contiguity is illustrated in Algorithm \ref{alg:pagetable_scanning}. On scanning each L2PTE, the 2MB virtual large page frame represented by the L2PTE is divided into 8 subregions. Then the L1PTEs of each subregion are scanned to judge whether each subregion is contiguous or not. Given a subregion, if any page of that subregion is not mapped contiguously with its adjacent pages in the same subregion, that subregion is discontiguous; otherwise, that subregion is contiguous and the corresponding $C_{x}$ bit in the L2PTE is set to 1. Besides, the contiguities between each adjacent subregions are also analyzed. If all subregions are contiguous and each subregion is also contiguous with its adjacent subregions in the same large page frame, then the $AC$ bit in the L2PTE is set to 1 to indicate the large page frame represented by the L2PTE is contiguous. When exploiting the contiguity of subregions, page permissions of each subregion are also considered, and subregions having pages with different permissions from other pages are regarded as discontiguous subregions. Besides, multiple contiguous subregions that can be coalesced together are also ensured to have the same permissions.

\textbf{Page Remapping, TLB Shootdown and MSC Eviction.}
Memory operations (e.g. memory deallocation and page migration) can change page mapping and existing memory contiguity. When updating L1PTEs of changed pages, the contiguity of corresponding subregions and large page frames are rechecked. Changing the mapping of one or a few pages can splinter a contiguous subregion or large page frame. In this case, in addition to updating L1PTEs of changed pages, the corresponding $C_{x}$ bits in L2PTE are also updated, and the $AC$ bit may also be updated. Besides, TLB shootdown is also triggered by OS to synchronize contiguity information. To reduce TLB shootdown overhead, only affected subregion TLB entries are evicted by setting the invalidation flag. Moreover, as changing the contiguity of just a single subregion may also alter the contiguity state between subregions in the same large page frame, the MSC is also checked to invalid entry of the changed large page frame. Remapping one or a few pages may also create a contiguous subregion or large page frame. In this case, L1PTEs of changed pages and contiguity bits in L2PTE are updated, and the regular TLB entries of changed pages are also flushed. Regular TLB entries of unchanged pages in the newly formed contiguous subregion or large page frame can continue to be used safely until evicted as the page mappings are not changed.

\section{Discussion and Future Work}
\label{sec:discussion}


\subsection{Combining with CoLT}
\label{sec:combine_colt}
The purpose of MESC is to exploit the existing advanced contiguity for GPUs. To augment the efficiency of basic MESC design, the existing intermediate contiguity can also be exploited by combining MESC with CoLT (MESC+CoLT). As discussed in previous sections, the support for MESC is only added in IOMMU TLB to keep per-CU TLBs simple and fast. In MESC+CoLT, the support for CoLT is added in per-CU TLBs. On walking the page table, regardless of whether the requested page resides in a contiguous subregion or not, the contiguity between pages (up to 4 pages in this paper) around the requested page is also checked and the coalesced translation of these pages is inserted into per-CU TLBs. Although,in this paper, we only combine MESC with CoLT, it should be noted that other  methods exploiting the intermediate contiguity can also be combined with MESC to further improve the benefits. 


\subsection{Optional Optimization for Discrete GPUs} 


Although the baseline model of this paper is based on GPUs integrated onto the same chips as CPUs, the ideas in this paper are also applicable to discrete GPUs having separate physical memory. Since the memory manager of discrete GPUs is mainly implemented in GPU driver/runtime and hardware, modifications, which is transparent to OS and application, to memory manager can be more flexible than integrated GPUs. As discussed in previous sections, the MSC is introduced to filter memory accesses when exploiting the contiguity between subregions in the same large page frame. Although MSC performs sufficiently well, however, in discrete GPUs, MSC can be totally removed by changing the layout of L1PTEs in page table pages pointed by L2PTEs as an alternative optimization. In the new L1PTE layout, the head L1PTEs of all 8 subregions in each virtual large page frame are placed in the beginning of the page table page sequentially. The benefit of the new L1PTE layout is that the up to 6 sequential memory accesses (explained in Section \ref{sec:page_table_walking}) to get the head L1PTEs of each contiguous subregion can be reduced or totally avoided when exploiting the contiguity between subregions. That is because the head L1PTEs of the 8 subregions occupy 64 bytes in total, which does not exceed the GPU cache line size \footnote{The GPU cache line is usually 128 bytes to accommodate memory requests from the active workitems in each wavefront.}. Thus all the 8 head L1PTEs can be read in burst without extra memory accesses when any of the 8 head L1PTEs is read during page table walk. To support the new L1PTE layout, the indexing mode of L1PTE should also be changed. As this method targets a different GPU architecture, we will evaluate this method in future work.

\section{Evaluation}
\label{sec:evaluation}

\subsection{Methodology}

\begin{table}[!t]
	\caption{The baseline system configuration.}
	\label{tab:baseline_configuration}
	\begin{tabular}{|l|m{5.5cm}|}
		\hline
		GPU					    & 700MHZ,16 CUs, 32 lanes per CU, 32 threads per wavefront					\\
		\hline
		L1 Data Cache 			& per-CU 32KB, 4-way associative, 128B cache line\\
		\hline
		L2 Data Cache 			& shared 2MB, 16-way associative, 128B cache line  \\
		\hline
		TLB 			        & 32-entry per-CU TLB, fully-associative\\
		\hline
		IOMMU		        	& 512-entry shared TLB, 16-way associative, 16 concurrent PTWs, 8KB PWC\\
		\hline
		DRAM				& 1GHZ, 2 channels, 8 banks per rank, FR-FCFS scheduler, burst length 8\\
		\hline
	\end{tabular}
\end{table}

\begin{figure*}
	\centering
	\includegraphics[width=7in]{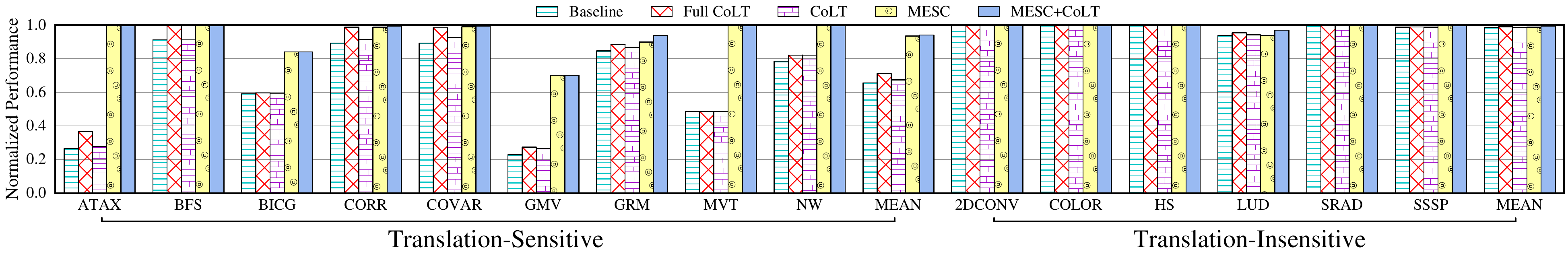}
	\caption{Performance for GPU workloads, normalized to the performance of THP.}
	\label{fig:eval_perf}
\end{figure*}

\textbf{Simulation.}
We evaluate MESC using a heterogeneous simulator gem5-gpu \cite{power2015gem5}, which models an integrated CPU-GPU system with a shared memory space. Table \ref{tab:baseline_configuration} shows the main parameters of the baseline system for our evaluations, including the GPU and the memory system. The OS support of MESC described in Section \ref{sec:ossupport} is implemented in Linux 3.4 to identify subregion contiguity of memory segments in heap. We also extend gem5-gpu to model the behavior of MESC and MESC+CoLT, including per-CU TLBs, unified shared TLB, PTW and MSC. 

\textbf{Workloads.}
We evaluate the performance of MESC with 15 workloads from various benchmark suites including Polybench \cite{pouchet2012polybench}, Rodinia \cite{che2009rodinia} and Pannotia \cite{che2013pannotia}. As discussed in previous section, all workloads are classified into 2 categories: \emph{translation-sensitive} (ATAX, BFS, BICG, CORR, COVAR, GMV, GRM, MVT and NW) and \emph{translation-insensitive} (2DCONV, COLOR, HS, LUD, SRAD and SSSP). To reduce simulation time, the input data is designed to enable each workload to be executed in one day. Although MESC mainly targets on translation-sensitive workloads, translation-insensitive workloads are also used to evaluate the overhead introduced by MESC. We use gem5-gpu's full-system mode to run workloads with our modified Linux kernel. 

\textbf{Comparison.}
We evaluate six separate implementations. (1) a \emph{baseline} architecture design with the state-of-the-art GPU address translation mechanism based on the work from Power et al. \cite{power2014supporting}; (2) \emph{thp}, an implementation that enabling 2MB transparent huge page \cite{arcangeli2010transparent}, which can be approximated as an ideal design with minimum address translation overhead; (3) \emph{full CoLT}, a design that can coalesce the translations of multiple contiguous base pages into per-CU TLBs and shared IOMMU TLB, which is similar to the coalescing mechanism proposed in \cite{pham2012colt} that has been implemented in AMD chips (e.g., Ryzen architecture); (4) \emph{CoLT}, a simplified variant of full CoLT that only inserts the coalesced translations to per-CU TLBs; (5) \emph{MESC}, a design that explores memory subregion contiguity, and we model a 512-entry MSC structure for MESC, which can filter most of the memory accesses in our evaluation when checking the contiguity between subregions; (6) \emph{MESC+CoLT}, an enhanced version of MESC that combines with the CoLT mechanism as discussed in Section \ref{sec:combine_colt}.

\textbf{Energy.}
We also report the dynamic energy spent in address translation path using CACTI 6.5 \cite{muralimanohar2007optimizing} with 32 nm process technology and the energy model proposed by Karakostas el al. \cite{karakostas2016energy}.


\begin{figure*}
	\centering
	\includegraphics[width=7in]{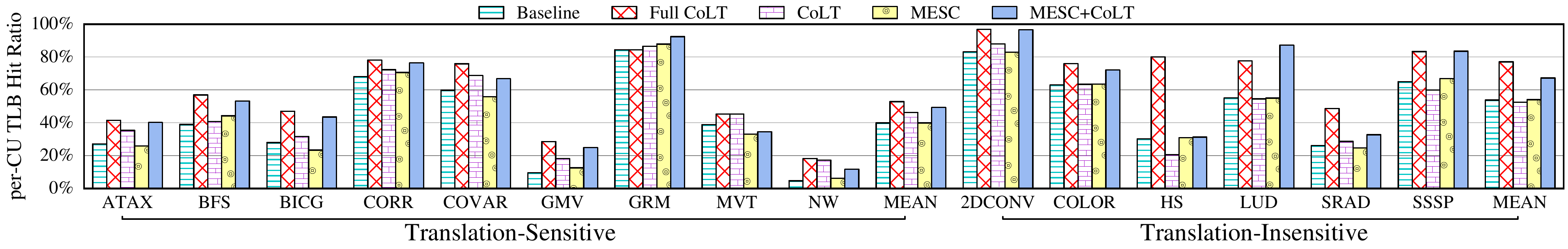}
	\caption{Per-CU TLB hit ratio for GPU workloads.}
	\label{fig:eval_l1tlb}
\end{figure*}

\begin{figure*}
	\centering
	\includegraphics[width=7in]{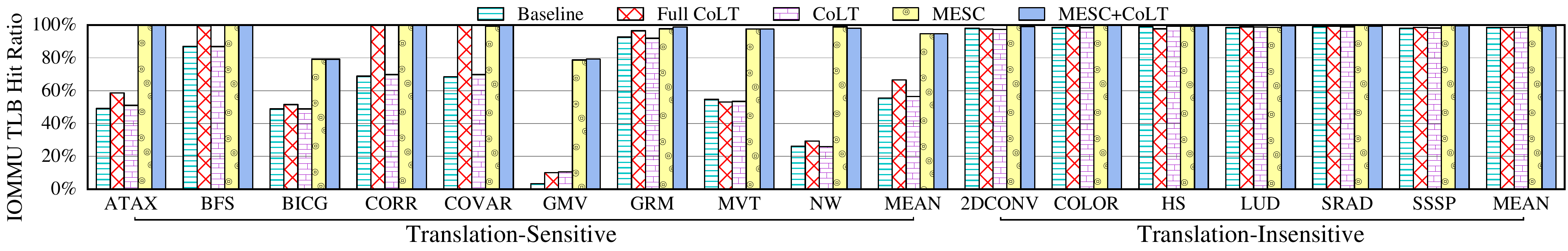}
	\caption{IOMMU TLB hit ratio for GPU workloads.}
	\label{fig:eval_l2tlb}
\end{figure*}

\subsection{Performance Analysis}
\label{sec:performance}

Figure \ref{fig:eval_perf} shows the performance of evaluated GPU workloads for baseline, full CoLT, CoLT, MESC and MESC+CoLT designs, normalized to the performance of THP. 
For translation-insensitive workloads, all the five implementations show similar results. That is because address translation is not the performance bottleneck for translation-insensitive workloads, and optimizing the address translation process will not improve the performance of these workloads obviously nor introduce extra overhead. 

For translation-sensitive workloads, the five designs produce different results. Since MESC exploits the advanced contiguity shown by the OS, which can cover considerable memory footprint, MESC can improve the performance to 93.5\% of the THP (i.e., 77.2\% performance improvement). In order to avoid increasing the access latency of per-CU TLBs, MESC only coalesces translations of contiguous memory subregions into IOMMU TLB. When MESC is combined with CoLT (MESC+CoLT) to coalesce translations of several consecutive pages into per-CU TLBs, the performance can be improved to 94.1\% of the THP (i.e., 78.0\% performance improvement). For most of the workloads (e.g., ATAX, MVT and NW), both MESC and MESC+CoLT almost eliminate the overhead of address translation. For some workloads (e.g., BICG and GMV), although there still exists significant address translation overhead even using MESC and MESC+CoLT, the performance of these workloads can also be greatly improved (e.g., the normalized performances of BICG and GMV using MESC are improved from 59.2\% to 84.1\% and from 22.8\% to 70.2\% respectively), which is much better than other designs. CoLT was designed to exploit the intermediate contiguity shown by the OS using a coalescing logic, which doesn't need to modify the OS. When coalesced translations are only inserted into per-CU TLBs, the performance of CoLT only reaches 67.4\% of the THP. When coalesced translations are inserted into both per-CU TLBs and IOMMU TLB, the performance of the full CoLT design can be improved to 71.1\% of the THP. Compared with other four designs, the performance of baseline design is the worst, which only reaches 65.5\% of the THP. From above results, we can see that coalescing translations of just several contiguous pages has very limited performance improvement for translation-sensitive workloads, and using the mechanisms proposed in this paper to coalesce translations of hundreds of contiguous pages can improve the performance of translation-sensitive workloads significantly, which illustrates the efficiency of MESC to reduce the overhead of address translation for GPUs.

\subsection{TLB Analysis}

The significant performance improvement for translation-sensitive workloads using MESC and MESC+CoLT is due to the fact that the coalesced translations enlarge the reach of TLBs, which improves TLB hit ratio greatly and filters most of the time-consuming page table walks. In this section we analysis the TLB hit ratio of each design and evaluate the sensitivity of MESC to the number of entries of different levels of TLB.

\textbf{TLB Hit Ratio.}
 Although MESC only coalesces translations into IOMMU TLB and keeps per-CU TLBs unchanged, the per-CU TLB hit ratios are also evaluated to show the influences made by Full CoLT, CoLT and MESC+CoLT designs. Figure \ref{fig:eval_l1tlb} shows the per-CU TLB hit ratios for the five designs. Because the per-CU TLB reach of MESC is not increased, the per-CU TLB hit ratios of MESC and baseline are at the same level. As full CoLT, CoLT and MESC+CoLT coalesce translations into per-CU TLBs, their per-CU TLB hit ratios are increased to different extents. For CoLT design, inserting coalesced translations into per-CU TLBs is only happened when both per-CU TLB and IOMMU TLB are missed. After the page table is walked, translation of the requested base page is inserted into IOMMU TLB and coalesced translation of several consecutive pages around the requested page is inserted into per-CU TLBs. If per-CU TLBs are missed and IOMMU TLB is hit, the base page translation will be inserted into per-CU TLBs, which reduces the per-CU TLB reach compared with the full CoLT design. Thus, the per-CU TLB hit ratio of CoLT is lower than full CoLT. For MESC+CoLT design, after the page table is walked, coalesced translation of contiguous memory subregions is inserted into IOMMU TLB and coalesced translation of several consecutive pages is also inserted into per-CU TLBs. Although the reach of IOMMU TLB in MESC+CoLT design is much larger than that in full CoLT design, the reach of per-CU TLB in MESC+CoLT design is not always larger due to that there are more base page TLB entries (e.g., at least half of the IOMMU TLB entries in MESC design and MESC+CoLT design are base page TLB entries). Thus, more base page TLB entries may be inserted into per-CU TLBs if per-CU TLBs are missed and IOMMU TLB is hit, which leads to that the per-CU TLB hit ratio in MESC+CoLT design is not always higher than full CoLT design. Inserting coalesced translation of several consecutive pages, which are not resident in a contiguous memory subregion, into IOMMU TLB indeed can improve both per-CU TLB hit ratio and IOMMU TLB hit ratio; however, that will further increase the complexity of hardware and the current MESC and MESC+CoLT designs perform sufficiently well.

Figure \ref{fig:eval_l2tlb} shows the IOMMU TLB hit ratios for the five designs. Because the IOMMU TLB hit ratios of translation-insensitive workloads in baseline design are high enough, the IOMMU TLB ratios are not changed significantly. For translation-sensitive workloads, the IOMMU TLB hit ratios in both MESC and MESC+CoLT designs are improved to nearly 95\%, which outperform other designs. The great improvement in IOMMU TLB hit ratio comes from coalescing translations of contiguous memory subregions into single IOMMU TLB entry, thus, each coalesced TLB entry can cover up to 512 base pages. In MESC, not only the translations of a contiguous large page frame can be coalesced, the translations of contiguous memory subregions in a discontiguous large page frame can also be coalesced, which increases the probability of TLB coalescing. CoLT design does not coalesce translations into IOMMU TLB, thus its IOMMU TLB hit ratio is not improved. Full CoLT design coalesces translations of only several contiguous pages into both IOMMU TLB and per-CU TLBs, and its IOMMU TLB hit ratio is improved to 66.5\%, which is much lower than MESC and MESC+CoLT.

\begin{figure}
	\centering
	\includegraphics[width=3.3in]{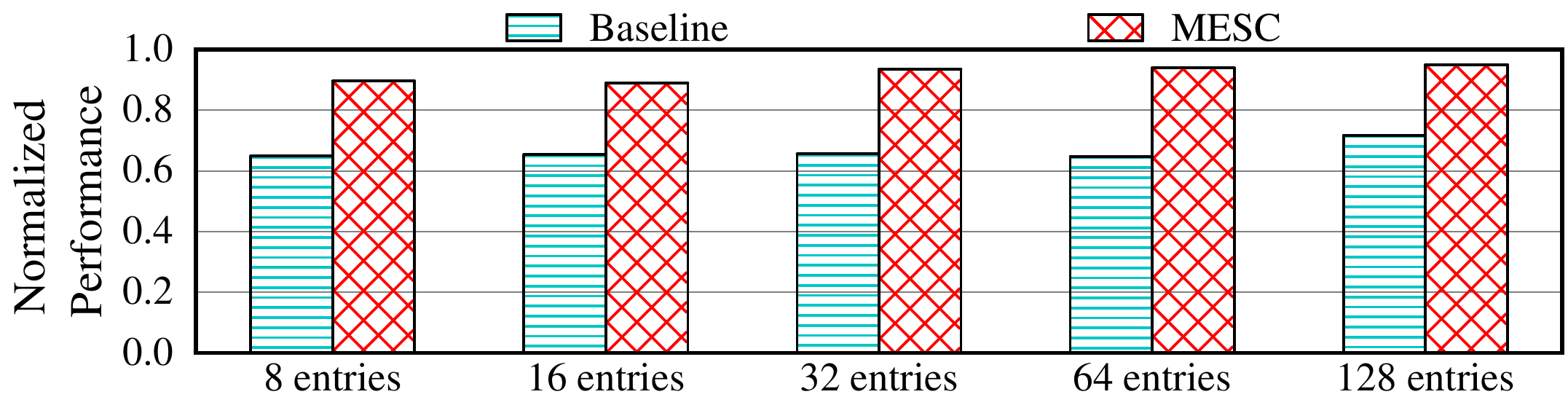}
	\caption{Performance for translation-sensitive GPU workloads across different per-CU TLB entries using MESC, normalized to the performance of THP. The number of IOMMU TLB entries is 512.}
	\label{fig:l1tlb_scalable}
\end{figure}

\begin{figure}
	\centering
	\includegraphics[width=3.3in]{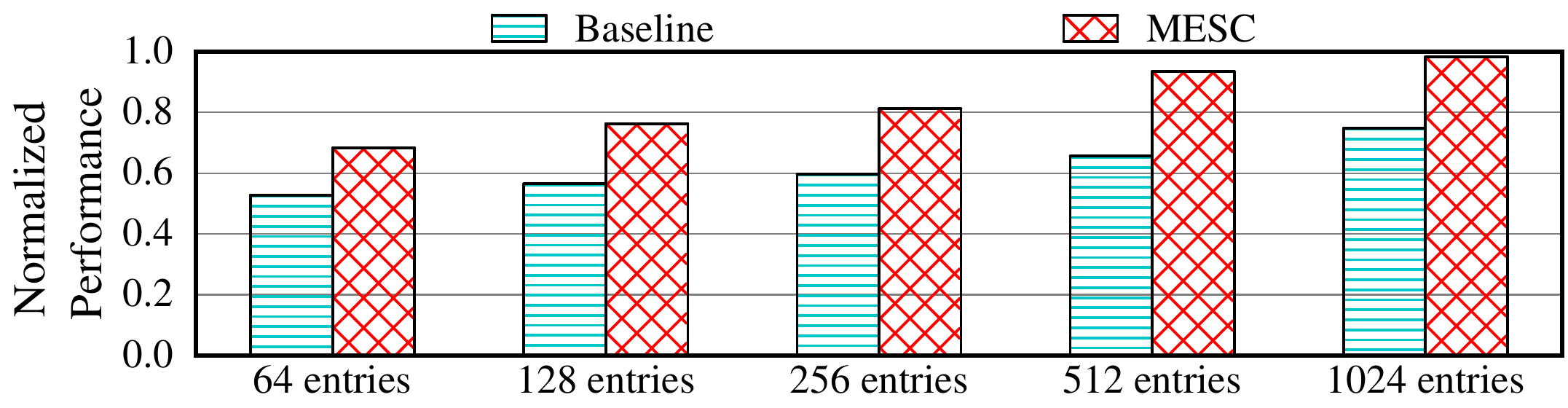}
	\caption{Performance for translation-sensitive GPU workloads across different IOMMU TLB entries using MESC, normalized to the performance of THP. The number of per-CU TLB entries is 32.}
	\label{fig:l2tlb_scalable}
\end{figure}

\textbf{TLB Sensitivity.}
MESC can improve TLB reach by coalescing translations of contiguous memory regions with various sizes into the IOMMU TLB. After coalescing, most of the address translations can hit in the IOMMU TLB, which reduces per-CU TLBs' pressure. By combining with CoLT, MESC+CoLT can improve the per-CU TLB hit ratio, but there is not much advantage in improving GPU performance (see Section \ref{sec:performance}). We now evaluate how sensitive the performance of MESC is to the number of entries in per-CU TLBs and IOMMU TLB.

Figure \ref{fig:l1tlb_scalable} and \ref{fig:l2tlb_scalable} show the performance for translation-sensitive GPU workloads across different per-CU TLB and IOMMU TLB entries using baseline and MESC respectively. All results are normalized to the performance of THP. From Figure \ref{fig:l1tlb_scalable} we can see that baseline is insensitive to the number of per-CU TLB entries when the number does not exceed 64. Even though the number of per-CU TLB entries is increased to 128, the normalized performance of baseline is only increased to 71.7\%, much lower than MESC. MESC is also insensitive to the number of per-CU TLB entries, but the difference is that even the number of per-CU TLB entries is only 8, MESC can obtain nearly 90\% performance of the THP. The reason is that MESC coalesces translations of contiguous memory subregions into IOMMU TLB, thus even though the per-CU TLB hit ratio is very low, most of the translation requests can hit in IOMMU TLB. From Figure \ref{fig:l2tlb_scalable} we can see that both baseline and MESC are sensitive to the number of IOMMU TLB entries. However, with the increase in the number of IOMMU TLB entries, the performance improvement of baseline is slow and the normalized performance is only 74.8\% even with 1024 IOMMU TLB entries. By contrast, the normalized performance of MESC reaches 81.2\% when the number of IOMMU TLB entries is only 256. Through the above analysis, we can say that MESC is suitable for accelerators especially that without or with very small private TLBs; besides, the capacity of IOMMU/shared TLB can also further be reduced, if the performance is acceptable, to save area and energy consumption using MESC.

\subsection{Address Translation Energy Consumption}

To report the dynamic energy consumption in address translation path, we use CACTI 6.5 \cite{muralimanohar2007optimizing} with the configuration in Table \ref{tab:baseline_configuration} to estimate the read energy and write energy for per-CU TLB, IOMMU TLB, MSC, PWC and DRAM. Because regular TLB entries and subregion TLB entries are looked up in different ways in MESC and MESC+CoLT, we consider the unified IOMMU TLB structure as 2 independent TLBs. One is a 512-entry, 16-way associative TLB for base pages, and another is a 256-entry, 8-way associative TLB for memory subregions, and the read energy and write energy of the two TLBs are estimated separately. Besides, we collect all per-CU TLB accesses, IOMMU TLB accesses, MSC accesses, PWC accesses and DRAM accesses during address translation. Then we use the energy model \cite{karakostas2016energy} to calculate all dynamic energy consumed by above structures in address translation path.

\begin{figure*}[htb]
	\centering
	\includegraphics[width=7in]{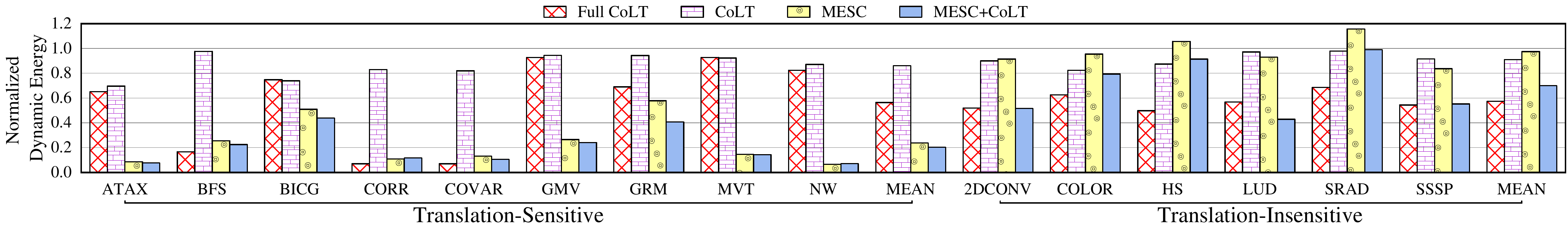}
	\caption{Dynamic energy spent in address translation path, normalized to the baseline.}
	\label{fig:energy_graph}
\end{figure*}

Figure \ref{fig:energy_graph} shows the dynamic energy consumptsion in address translation path, normalized to the baseline design. By coalescing the translations of hundreds of contiguous pages, our designs can improve TLB hit ratio significantly and filter most of the memory accesses in address translation. Our designs can not only improve performance (see Section \ref{sec:performance}) greatly, but also reduce dynamic energy consumption in address translation path significantly. The energy benefits are obvious especially for translation-sensitive workloads, this is because the TLB hit raios of translation-sensitive workloads are very low and the memory is accessed frequently in page table walking. From Figure \ref{fig:energy_graph} we can observe that MESC and MESC+CoLT obtain a reduction of 76.4\% and 79.7\%, respectively, in dynamic energy consumption for translation-sensitive workloads.
Full CoLT and CoLT only coalesce several contiguous pages, their dynamic energy reduction for translation-sensitive workloads are 43.6\% and 14\% respectively.

For translation-insensitive workloads, MESC design only shows a 2.5\% reduction in dynamic energy consumption. Although MESC can eliminate most of the memory accesses in address translation and reduce the dynamic energy consumption in address translation path, there also exist other energy consumptions compared with baseline in MESC such as MSC lookups and extra memory accesses to explore the contiguity between subregions. By combining with CoLT to insert coalesced translations into per-CU TLBs, which improves the per-CU TLB hit ratio and reduces the IOMMU TLB lookups, MESC+CoLT design shows a 30\% reduction in dynamic energy consumption for translation-insensitive workloads. We also note that coalescing translations of contiguous pages resident in discontiguous subregions into IOMMU TLB can further reduce energy consumption. We leave this exploration to future work.

\subsection{Fragmentation analysis}
To evaluate the effect of fragmentation to MESC, we run the memhog tool provided by numactl \cite{wickman2015numactl} to fragment the memory randomly. The kernel \emph{defrag} flag is also set up to show the effect of memory compaction. We evaluate the percentage of memory footprint of applications that can be covered by contiguous subregions, which can further be exploited by MESC, under the situations that 25\%, 50\% and 75\% of the memory is fragmented respectively. 

\begin{table}[h]
	\centering
	\caption{Percentage of memory footprint that is covered by contiguous subregions with varying memhog.}
	\label{tab:memhog_analysis}
	\begin{tabular}{|l|p{1.2cm}<{\centering}|p{1.2cm}<{\centering}|p{1.2cm}<{\centering}|}
		\hline
		\diagbox[width=12em]{Defrag flag}{Subregion}{Memhog}	& 25\% & 50\% &  75\% \\
		\hline
		Enable	& 48.7\% & 42.8\% & 38.9\% \\
		\hline
		Disable	& 44.3\% &  42.3\% & 34.7\%	\\
		\hline
	\end{tabular}
\end{table}

The results are shown in Table \ref{tab:memhog_analysis}. We can observe that more contiguous subregions can be allocated when the defrag flag is enabled; besides, more than 34.71\% of the total memory footprint can be exploited by MESC to enlarge TLB reach, even when the memory is heavily fragmented (75\%) with the defrag flag disabled. That is because address translations are coalesced in granularity of subregion, and MESC is still applicable in high load system with little large contiguous memory regions.

\section{Related Work}
\label{sec:related_works}

To improve the efficiency of address translation, modern processors and OSes \cite{arcangeli2010transparent,mel2010libhugetlbfs} support multiple page sizes. As large pages can increase the coverage of TLB (e.g., each 2MB page TLB entry can cover 512 base pages), TLB hit ratio can be improved significantly. To support multiple page sizes, processors usually maintain separate TLBs for each page size (e.g., Intel Sandybridge, Haswell and Skylake architectures maintain separate L1 TLBs for 4KB, 2MB and 1GB page sizes \cite{intelhaswell,intelskylake,cox2017efficient}). To improve utilization of separate TLBs, Cox el al. \cite{cox2017efficient} propose MIX TLBs that accommodates multiple page sizes in the same TLB. However, MIX TLBs relies on the contiguity of consecutive large pages. The unified TLB in this paper also supports multiple page sizes, and the difference is that our unified TLB mainly targets the internal contiguity of virtual large page frames. In some architectures, multiple page sizes can also be accommodated in the same fully-associative TLB. For example, in ARM architecture \cite{armmaintlb}, each main TLB entry contains a page size field to indicate different page sizes, but the fully-associative TLB determines the number of TLB entries is limited.

TLB coalescing is another important technique to reduce the overhead of address translation. TLB coalescing is mainly to increase the reach of TLB by coalescing translations of consecutive pages in a single TLB entry. Direct Segments \cite{basu2013efficient} and RMM \cite{karakostas2015redundant} can coalesce contiguous memory regions with unlimited sizes. However, Direct Segments only support one single contiguous region for each application. Although RMM can support multiple contiguous regions, it requires the OS to allocate a few contiguous regions as large as possible; besides it needs large and
power-hungry Range TLBs, which may be prohibitive given
the area and power budgets of accelerators \cite{haria2018devirtualizing}. Above methods require to modify the default memory allocation policy in OS (e.g., using eager paging) to allocate contiguous memory regions. Eager paging is not necessary in Hybrid TLB \cite{park2017hybrid}, but the TLB shootdowns can severely hurt performance. Other works (such as CoLT \cite{pham2012colt}, Sub-block TLB \cite{talluri1994surpassing} and Cluster TLB \cite{pham2014increasing}) coalesce translations of multiple pages with none or less OS modifications, but they usually only coalesce very limited translations (e.g., 4-8). MESC can coalesce 64-512 translations in a single TLB entry. Mosaic \cite{ausavarungnirun2017mosaic} makes a trade-off between TLB reach and demand paging performance for discrete GPUs, but the internal contiguity of large page frames, which is exploited in MESC, is not considered.

How to reduce TLB miss penalty is also a hot topic to improve the efficiency of address translation. 
Shin et al. \cite{shin2018scheduling} discover that the order of servicing page table walk requests plays an important role in improving GPU performance for irregular applications. They propose a novel page table walk scheduling mechanism combining shortest-job-first (SJF) and bach policies and improve the performance of irregular applications significantly. Virtual caching \cite{wood1986cache,basu2012reducing,kaxiras2013new,park2016efficient,yoon2016revisiting} has been proposed for a long time for CPUs to reduce the latency and energy of address translation as data can be read directly from cache using virtual addresses. Yoon et al. \cite{yoon2018filtering} observe that many translation requests that miss in GPU TLBs can obtain corresponding data, which can be read using the translated physical addresses, in the GPU cache hierarchy. Thus, to avoid the unnecessary TLB lookups and page table walks, they apply virtual caching to GPUs. However, the problems of synonyms, multi-processing and cache coherence \cite{park2016efficient,yoon2016revisiting,kaxiras2013new,yoon2018filtering} are complex to implement.
\section{Conclusion}
\label{sec:conclusion}

Address translation, which is the key factor of virtual memory, is becoming the bottleneck of GPU performance. To reduce the overhead of address translation, the TLB hit ratio should be improved to reduce the frequency of time-consuming page table walks. In this work, we identify the opportunity of advanced memory contiguity presented by the OS and introduce MESC to coalesce the translations of contiguous memory regions with various sizes. The key idea of MESC is to divide each large page frame in virtual memory space into subregions with fixed size , and then store the contiguity information of each subregion and large page frame in L2PTEs; besides, the contiguity information between subregions in discontiguous large page frames is cached in MSC. With above mechanisms, address translations of up to 512 base pages can be coalesced into single TLB entry, which can improve the TLB reach significantly, without the needs of changing the default memory allocation policy and the support of large pages. Through the evaluation, we have shown that MESC achieves significantly performance improvement and dynamic energy reduction in address translation path for translation-sensitive workloads.
\ifCLASSOPTIONcompsoc
\section*{Acknowledgments}
\else
\section*{Acknowledgment}
\fi

This work is supported by the National Key Research and Development Program of China under Grant No. 2016YFB1000503 and the National Science Foundation of China under Grant No. 61572062.

	
\ifCLASSOPTIONcaptionsoff
\newpage
\fi


%



\bibliographystyle{IEEEtran}
\bibliography{IEEEabrv,sections/main}
\end{document}